# Entanglement and control of single quantum memories in isotopically engineered silicon carbide

Alexandre Bourassa[1,†], Christopher P. Anderson[1,2,†], Kevin C. Miao[1], Mykyta Onizhuk[1,3], He Ma[1], Alexander L. Crook[1,2], Hiroshi Abe[4], Jawad Ul-Hassan[5], Takeshi Ohshima[4], Nguyen T. Son[5], Giulia Galli[1,3,6], David D. Awschalom[1,2,6,*]

[1] *Pritzker School of Molecular Engineering, University of Chicago, Chicago, IL 60637, USA*
[2] *Department of Physics, University of Chicago, Chicago, IL 60637, USA*
[3] *Department of Chemistry, University of Chicago, Chicago, IL 60637, USA*
[4] *National Institutes for Quantum and Radiological Science and Technology, 1233 Watanuki, Takasaki, Gunma 370-1292, Japan*
[5] *Department of Physics, Chemistry and Biology, Linköping University, SE-581 83 Linköping, Sweden*
[6] *Center for Molecular Engineering and Materials Science Division, Argonne National Laboratory, Lemont, IL 60439, USA*
\* *Corresponding author*
† *These authors contributed equally to this work*

**Nuclear spins in the solid state are both a cause of decoherence and a valuable resource for spin qubits. In this work, we demonstrate control of isolated $^{29}$Si nuclear spins in silicon carbide (SiC) to create an entangled state between an optically active divacancy spin and a strongly coupled nuclear register. We then show how isotopic engineering of SiC unlocks control of single weakly coupled nuclear spins and present an *ab initio* method to predict the optimal isotopic fraction which maximizes the number of usable nuclear memories. We bolster these results by reporting high-fidelity electron spin control (F=99.984(1)%), alongside extended coherence times ($T_2$=2.3 ms, $T_2^{DD}$>14.5 ms), and a >40 fold increase in dephasing time ($T_2^*$) from isotopic purification. Overall, this work underlines the importance of controlling the nuclear environment in solid-state systems and provides milestone demonstrations that link single photon emitters with nuclear memories in an industrially scalable material.**

Nuclear spins are one of the most robust quantum systems, displaying relaxation times that can exceed hours or days[1–3]. This makes them exciting candidates for quantum technologies requiring long memory times. In particular, nuclear spins are attractive quantum registers for optically active spin defects in the solid-state[4]. For example, nuclear registers can be used for repetitive quantum non-demolition (QND) optical readout[5], to enhance the signal-to-noise in quantum sensing[6], to implement quantum error correction schemes[7], or as vital components of quantum repeater[8] and quantum communications[9] nodes. Additionally, electron-nuclear hybrid systems provide a platform for studying measurement back-action[10] and the emergence of classicality in quantum mechanics[11].

Recently, commercial SiC has been shown to provide a technologically mature semiconductor host for multiple defect spin qubits[12–17]. In particular, this material allows the integration of isolated color centers into classical electronic devices which can be used to engineer and tune the spin-photon interface[18]. Combining such a tunable near-infrared emitter[19,20] with a long-lived quantum memory is a promising basis for quantum network nodes fabricated at wafer scale by the semiconductor industry. To realize these quantum memories, SiC provides both carbon and silicon isotopes with non-zero nuclear spin. These isotopes have been shown to couple to various electronic spin defects[12,21]; however, the control of single nuclear spins[22] in SiC has remained an outstanding challenge.

In this work, we report coherent control and entanglement of nuclear spin quantum registers strongly coupled to a single neutral divacancy spin ($VV^0$) in naturally abundant SiC. We then extend this control to weakly coupled nuclear spins, where isotopic purification enables the isolation of robust quantum memories. Using isotopic engineering, we also report both record coherence times and record single qubit gate fidelities[23] for electronic spins in SiC. Throughout this work, we present both experiment and *ab initio* theory that explores the inherent tradeoffs between spin coherence and nuclear memory availability which are involved when isotopically engineering materials. These results develop a full suite of nuclear spin controls in SiC and provide a guide for future materials design of spin-based quantum technologies.

**Strongly coupled nuclear registers**

In natural SiC, 1.1% of the carbon atoms and 4.7% of silicon atoms possess an $I = 1/2$ nuclear spin. Thus, about a third of all single *c*-axis oriented (*hh* and *kk*, Supplemental Information) divacancies will have a $^{29}$Si register on one of the nearest-neighbor lattice sites (denoted $Si_I$, $Si_{IIa}$ or $Si_{IIb}$)[24]. When the hyperfine coupling exceeds the linewidth (order $1/T_2^*$) of the electronic state (Fig. 1a), oscillations due to these nuclear spins are observable in Ramsey experiments. We refer to such nuclear spins as strongly coupled. This strong coupling splits the $m_s = \pm 1$ electronic ground state levels, which results in pairs of resolved transitions that enables direct selective control of this two-qubit state using external radio frequency (RF) magnetic fields.

Here, we demonstrate such a strongly coupled system by isolating a single *c*-axis (*kk*) VV$^0$ with a nearby $^{29}$Si at the $Si_{IIa}$ site (parallel hyperfine $A_{||} = 2\pi \cdot 13.2$ MHz) in natural 4H-SiC. In this case, because the electron spin linewidth (~1 MHz) is much lower than the hyperfine splitting $A_{||}$, we observe two individually addressable transitions corresponding to the two nuclear spin states (Fig. 1b). To polarize this nuclear register, we make use of two iterations of algorithmic cooling in which we optically polarize the electron spin and then swap this polarized state to the nuclear spin[25]. Using this method, we can achieve a high initialization fidelity (~93%) as measured by the peak asymmetry in the optically detected magnetic resonance spectrum shown in Fig. 1b (Supplemental Information).

After nuclear initialization, we prepare the electron spin in the $m_s = -1$ state and use a 13.2 MHz RF magnetic field to drive nuclear Rabi oscillations (Fig. 1c), which we read out by projecting onto the electron spin. Since these oscillations are only driven in the $m_s = \pm 1$ states, this allows us to demonstrate a $C_{\pm 1}NOT_n$ gate[12] which can be performed in 12.7 μs. Throughout these measurements, we also make use of fast (limited only by the hyperfine splitting of the lines) $C_nNOT_e$ gates by applying microwave pulses at one of the two frequencies shown in Fig. 1d.

Having demonstrated control over a single nuclear spin, we then increase the number of registers by finding a (*kk*) divacancy which is strongly coupled to two $^{29}$Si spins (with 6% probability for naturally abundant SiC). For this defect, we show that by using both algorithmic cooling and dynamical nuclear polarization[12,26] (DNP), we can polarize the full three-qubit system (Fig. 1b). We then demonstrate individual control of these registers and calibrate gates operating on either register (Supplemental Information).

In this three-qubit spin system, we apply the quantum circuit in Fig. 1e on the electron and one of the two coupled nuclear spins to create an electron-nuclear entangled state, and measure its full density matrix using quantum state tomography[12] (QST). We evaluate this density matrix using the positive partial transpose test, confirming unambiguously the entanglement in this system with an estimated entangled state fidelity of ~81% (Supplemental Information).

These results demonstrate that single, strongly coupled nuclear spins can be used as quantum registers in SiC with relatively fast gate times. This type of register is useful for QND measurement of the nuclear spin and more generally for any applications that require fast operations[27] on ancilla qubits[28,29]. However, the number of available nearby nuclear sites which can be controlled in this way is limited. Additionally, the high coupling strength makes these nuclear registers more sensitive to stochastic noise from the electron spin and limit applications where repeated electron initialization and control is necessary[8,30], such as in long-distance quantum communications[31] or entanglement distillation[9].

**Weakly coupled nuclear memories**

To complement these strongly coupled registers, we therefore investigate nuclear spins which are weakly coupled to divacancy electron spins. In order to access these memories and go beyond the $1/T_2^*$ limit, we use an XY8-based dynamical decoupling sequence to perform nanoscale NMR[22,32–34] of the nuclear environment of a (*kk*) divacancy (Fig. 2a). This sequence (Fig. 2b) not only protects the electron spin from decoherence, but also allows for selective control of nuclear spins even when their hyperfine coupling is lower than the electron spin linewidth. In this measurement, each nuclear spin produces a series of dips in the coherence function at a pulse spacings[22] $\tau_k \approx \frac{(2k+1)\pi}{2\omega_L + A_{||}}$ at integer order $k$ and Larmor frequency $\omega_L$, corresponding to its specific nuclear precession frequency. With this spectroscopy, we observe that natural SiC has a crowded nuclear resonance spectrum due to the relatively abundant $^{29}$Si, making it difficult to isolate single spins with low hyperfine coupling[30] (defined here to be < $2\pi \cdot 60$ kHz). This spectrum, along with *ab initio* cluster-correlation-expansion[35] (CCE) simulations of various possible nuclear spin configurations (Fig. 2a), demonstrates that natural SiC is not well suited for isolating single weakly coupled nuclear spins with low hyperfine values.

To address this issue, we use isotopically purified gases to grow 4H-SiC with 99.85% $^{28}$Si and 99.98% $^{12}$C (Methods). In this sample, we once again measure the nuclear environment of a few (*kk*) divacancies and identify one with a single isolated dip in the coherence function (Fig. 2a). We find that the dip positions very closely match the different orders (*k*) of the Larmor frequency of a $^{29}$Si (differing only through the hyperfine value[22]). We further confirm the gyromagnetic ratio for this nuclear spin species by repeating the experiment at a different magnetic field (Supplemental Information).

Having confirmed that the dips correspond to a $^{29}$Si nuclear spin, we perform spectroscopy in both the {$m_s = 0$, $m_s = +1$} and the {$m_s = 0$, $m_s = -1$} basis (Fig. 2c), and measure a small $A_{||} \approx 2\pi \cdot 650$ Hz[22], which would not be resolvable in a Ramsey experiment. Low $A_{||}$ nuclear spins are especially useful as robust quantum memories because the dephasing of the nuclear spin caused by stochastic noise from the electron is particularly sensitive to the parallel component of the hyperfine tensor ($A_{||}$)[8].

Fixing the pulse spacing (2τ) to a specific coherence dip ($k = 6$), we then vary the number of pulses (*N*) to coherently control this weakly coupled single nuclear spin[7,22]. The corresponding $C_eROT_{x,n}(\pm\theta)$ oscillations observed (Fig. 2d) allow us to measure the perpendicular hyperfine component $A_\perp \approx 2\pi \cdot 11.45$ kHz (where $\theta \approx \frac{A_\perp \cdot N}{\omega_L}$) and confirm the successful application of a maximally entangling two-qubit gate[7] (Supplemental Information). If no other nuclear spins were present, one could choose any resonance order (*k*) to perform the two-qubit gate. In practice however, as *k* increases, the resonance of the isolated nuclear spin separates from the rest of the bath which drastically increases the two-qubit gate fidelity. Here, even in the isotopically purified sample where the nuclear spectrum is sparse, the electron-nuclear gate fidelity increases greatly at higher orders ($k$) as the resonance separates from the bath (up to 97(1)% at $k$ = 6, Supplemental Information). These results demonstrate the importance of reducing the nuclear spin bath for high fidelity control of isolated quantum memories with weak hyperfine interactions.

With these results in mind, we now turn our attention to estimating the optimal isotopic fraction required to maximize the number of isolated and controllable nuclear memories. Here, we need to strike a balance between too much purification which removes most usable nuclear spins and too little which results in a crowded and unresolvable spectrum. Limiting the gate time to a regime where nuclear-nuclear interactions are negligible (Supplemental Information), we developed a method to predict the average number of resolvable nuclear memories as a function of isotopic concentration. This is achieved by considering both the intrinsic gate fidelity from the electron-nuclear interaction and the average effect of unwanted rotations from all other nuclear species (Methods). Our analysis demonstrates several important aspects of nuclear availability in SiC.

First, there exists an optimal nuclear spin concentration (Fig. 3a) that maximizes the average number of available nuclear memories which can be controlled within a maximum gate time and at a given minimum gate fidelity. Here, we find that naturally abundant SiC has a prohibitively high concentration of $^{29}$Si, which prevents the isolation of nuclear memories with low hyperfine coupling (<2π·60 kHz). This reinforces the importance of isotopic engineering for nuclear memories in SiC and explains the spectrum observed in Fig. 2a. Second, the hyperfine values of the resulting controllable memories vary with isotopic concentration (Fig. 3b). At high concentration, nuclei with moderate hyperfine (>2π·60 kHz) contribute to most of the available memories, while low hyperfine nuclear spins are unresolvable. On the other hand, a lower isotopic concentration results in a less crowded spectrum and allows for the isolation of nuclei with lower hyperfine. The choice of nuclear concentration thus not only determines the total number of available quantum memories, but also the distribution of hyperfine values for these controllable nuclei.

Furthermore, we note that there is a tradeoff between the maximum allowable gate time and the number of available nuclear memories. While longer gate times allow for the resolution of more distant nuclei, this increase is shown to be only sublinear (Supplemental Information). Additionally, when both nuclear species are utilized, the SiC binary lattice may provide roughly double the number of resolvable nuclear registers compared to a monoatomic crystal.

While the range of desired hyperfine values may differ depending on the particular application, a careful selection of the isotopic fraction is critical to maximizing the number of nuclear spins available in this range. This careful selection also determines the resulting average gate speeds and fidelities, allowing further optimization for the application at hand. These results therefore constitute not only a proof-of-principle demonstration of single weakly coupled nuclear spin control in SiC, but also provide guidance for future isotopic growth of materials for a variety of spin-based quantum technologies.

**High-fidelity qubit control and extended coherences**

Broadly, these experiments are all predicated on the divacancy electronic spin being a controllable and long-lived qubit. In this section, we discuss in detail the main factors that limit the coherence of divacancies in SiC and quantify our ability to perform single-qubit manipulation.

We begin by measuring both $T_2^*$ (Ramsey spin dephasing time) and $T_2$ (Hahn-echo coherence time) of both *c*-axis (*kk*) and basally (*kh*) oriented defects in isotopically purified material. We measure the *c*-axis defects at B=48.8 G and the basal defects at B=0 G (to benefit from the magnetic insensitivity arising from a clock-like transition[36,37]).

We report (Fig. 4a and 4b) $T_2^*$ times of 48.4(7) μs and 375(12) μs for the *c*-axis (*kk*) and basal (*kh*) defects in isotopically purified SiC, compared to 1.1 μs[38] and 70-160 μs[36,37] in naturally abundant material. These numbers correspond to record dephasing times for spin qubits in SiC[21]. Additionally, despite only moderate isotopic purity, these results are very competitive with NV centers in diamond with much lower nuclear spin concentration[39–41]. This favorable scaling most likely arises from the SiC binary lattice and longer bond length, which results in reduced nuclear flip-flops[35]. These improvements in $T_2^*$ are vital for DC quantum sensing schemes and for achieving strong coupling in hybrid systems[42,43].

The significant increase in dephasing times arising from the isotopic purification for the *c*-axis defects shows that magnetic field noise from the nuclear environment is by far the main limiting factor to $T_2^*$ for these defects. We provide further evidence of this by investigating the dephasing in isotopically purified SiC with *ab initio* cluster-correlation-expansion (CCE) simulations. Taking into account the remaining nuclear spin bath, these calculations predict average $T_2^*$ values which are consistent with our experimental observations (Fig. 4a).

On the other hand, while basal divacancies benefit from first-order insensitivity to magnetic field noise at B=0 G, this magnetic noise protection comes at the cost of increased sensitivity to electrical fields[44]. Since charge fluctuations can cause significant electric field noise[18], this may explain why the increase in $T_2^*$ obtained from isotopic purification (Fig. 4b) is less pronounced than that of the *c*-axis divacancies. Furthermore, this magnetic protection also makes nuclear control difficult in the basal (*kh*) divacancies. This underlines the tradeoffs involved when choosing a defect species to work with.

Next, we perform Hahn-echo experiments to measure $T_2$ in isotopically purified SiC (Fig. 4c). Although we find a factor of ~2 improvement in the coherence time for (*kk*) defects in this material (2.32(3) ms versus 1.1 ms[38]), we remark that this is a more modest improvement than that of $T_2^*$. Nevertheless, this $T_2$ is comparable to the longest observed Hahn-echo coherence time in isotopically purified diamond samples with much greater isotopic purity[45,46]. Interestingly, the measured $T_2$ deviates from the predictions of nuclear spin induced decoherence obtained with CCE calculations, which yield an average coherence time of ~37 ms. To understand these results, we carried out second order CCE simulations to study the effect of non-interacting electron spin pairs on the coherence time[47]. At the estimated paramagnetic density (impurities and radiation induced defects in the $3\times10^{14}$-$3\times10^{15}$ cm$^{-3}$ range, Methods) we find good agreement with experiment (Fig. 4d), thus confirming both the accuracy of our theoretical model and the important role of paramagnetic defects in limiting coherence.

Our results are consistent with magnetic noise from a weak, but quickly fluctuating paramagnetic spin bath combined with noise from a strong, but slowly fluctuating, nuclear spin bath[48]. As a consequence, $T_2^*$ is limited by nuclear spins, while $T_2$ is limited by paramagnetic impurities for the *c*-axis defects. On the other hand, differences in the basal divacancy's coherence compared to other reports[36,37] likely stems not only from the isotopic purification, but also from sample-to-sample variations in electric field noise, which could be mitigated using charge depletion techniques[18].

The demonstrated coherence can be further extended by additional refocusing pulses. We provide a proof-of-principle demonstration by varying the number of pulses (*N*) in a dynamical decoupling sequence. At $N = 32$, the coherence is increased to 14.5 ms (for a *kk* defect, Fig. 4e). With more pulses, the coherence should continue to increase linearly until the $T_1$ limit is reached, which we measure to be on the order of one second under these experimental conditions (Supplemental Information).

Finally, we characterize our single qubit gate fidelities through randomized benchmarking experiments and obtain an average gate fidelity of 99.984(1)% (Fig. 5). These bare fidelities are amongst the highest for single spins in the solid state[23,49,50] and exceed the threshold for error correction codes[51–53]. Furthermore, high-fidelity control of the electron spin is crucial to prevent reduced coherence in nuclear spin memories[30]. The long coherence ($T_2^{DD} > 14.5$ ms) and high-fidelity control (99.984(1)%), combined with a >99% initialization and readout fidelity (Supplemental Information) demonstrated in this work establishes the divacancy in SiC as a promising system for future solid-state quantum devices.

**Conclusion**

Defect spins in SiC are exciting candidates for wafer-scale quantum technologies requiring stationary qubits and a photonic quantum communication channel. In this work, we provide milestone demonstrations of nuclear memory control of both strongly and weakly coupled nuclear memories in a technologically mature semiconductor material. This work also examines, both experimentally and theoretically, the tradeoffs that are inherent to isotopic purification and offers a pathway towards optimizing nuclear spin concentration to maximize the number of usable nuclear memories.

Our results underline the importance of isotopic engineering in designing materials for solid-state quantum applications. Such engineering can provide a two-fold benefit for quantum memories: it enables control of more nuclear spins by unlocking access to memories with low hyperfine coupling, while also drastically increasing the coherence of these nuclear spins[54]. Moreover, isotopic engineering enables the selection of a hyperfine distribution that can optimally trade off the effect of the "frozen core"[55] against the electron spin induced noise inherent in realistic quantum communications protocols[30]. Further optimization may also be achieved by considering differing nuclear control methods[4,56]. Additionally, the demonstrated proof-of-principle nanoscale NMR detection of a single nuclear spin (at a distance of ~1.2 nm) in SiC provides a route for a functionalizable, biocompatible platform for quantum sensing with polarization and readout in the biological near-infrared window[57]. Overall, these results cement defects in SiC as attractive systems for the development of quantum communication nodes and underline the importance of isotopic control in material design for future quantum technologies.

**Methods**

*Single defect observation and control*

Single defects are observed in a home-built confocal microscope operating at T= 5 K with a Montana Cryostation s100 closed-cycle cryostat. We utilize a high NA (0.85) NIR objective and single-mode fiber coupled (1060XP) IR-optimized SNSPD (Quantum Opus) and observe single defects with 40-50 kcts at saturation. 905 nm excitation is used along with a weak 705 nm tone for charge stabilization[18]. Microwave striplines are fabricated alongside an electrical control planar capacitor (10 nm Ti, 150 nm Au) using electron beam lithography. In 4H-SiC, single PL1 (hh), PL2 (kk) and PL4 (kh) defects are observed and are labelled following the $V_cV_{Si}$ convention and where *h* represents the hexagonal lattice site and *k* the quasi-cubic lattice site. The *c*-axis refers to the crystallographic axis in SiC which corresponds to the stacking direction of the hexagonal layers of SiC ([0001]). Basal defects are oriented along one of the basal planes. Resonant readout and initialization[18,19] (realized using a tunable Toptica DLC

PRO laser) can result in Rabi contrast exceeding 99% in optically detected magnetic resonance (ODMR) (Supplemental Information). This corresponds to the highest Rabi contrast reported in SiC and provides an achievable lower bound for initialization and readout errors combined. Reported coherences are for representative single defects.

For the strongly coupled nuclear spin experiments, Gaussian pulse shaping is used to perform spectrally narrow manipulation of the quantum registers. $^{13}$C registers are also available[24], but occur with lower probability in both the natural and isotopic samples. For nuclear spin spectroscopy and control, randomized benchmarking and coherence measurements, square pulses were used with π pulse times ranging from 50 ns to 1 μs. Magnetic fields are applied with a large permanent magnet on a goniometer. Alignment at high field is achieved by reducing the mixing from off axis magnetic fields, visible through the PLE magnitude after initializing the spin (a measure of cyclicity). In order to zero the magnetic field for kh divacancies, we utilize a three-axis electromagnet. Using a nearby c-axis *kk* defect as a magnetometer, the field is zeroed by reducing the splitting between the very narrow CW ODMR lines in the isotopically purified sample (<20 kHz).

*Materials growth*

Natural 4H-SiC was obtained from Norstel AB (now ST Microelectronics) in the form of a 20 μm intrinsic epitaxial layer grown on 4° off-axis HPSI 4H-SiC. This layer contains <1×10$^{15}$ cm$^{-3}$ $V_c$. For the isotopically purified ("isotopic") sample, epitaxial 4H-SiC was CVD grown on a 4° off-axis n-type 4H-SiC substrate at a thickness of ~90 μm using isotopically purified Si and C precursor gasses. The purity is estimated to be 99.85% $^{28}$Si and 99.98% $^{12}$C, which was confirmed by secondary ion mass spectroscopy (SIMS). C-V measurements show a slightly n-type behavior of this layer with a free carrier concentration of 6×10$^{13}$ cm$^{-3}$. This roughly matches the measured concentration (3.5×10$^{13}$ cm$^{-3}$) of nitrogen through comparisons of the bound exciton lines. DLTS places the $V_c$ concentration at the mid 10$^{12}$ cm$^{-3}$ range before irradiation.

In the naturally abundant material, single defects are created using a 1×10$^{13}$ cm$^{-2}$ dose of 2 MeV relativistic electrons. Subsequent annealing at 810°C in an Ar environment produces spatially isolated single VV$^0$. For the isotopically purified material, an electron dose of 1×10$^{13}$ cm$^{-2}$ (Fig. 5) and 5×10$^{14}$ cm$^{-2}$ (Fig. 2-4) are used. Despite the low impurity and defect content of the starting material, this means that the number of induced displacements[58] in the lattice after irradiation can be as high as 0.5-3 cm$^{-1}$ × (dose)=(0.25-1.5)×10$^{15}$ cm$^{-3}$. These defects can be paramagnetic and most likely consists of $V_C$, $V_{Si}$ and associated vacancy complexes. This is also a relatively common range even before irradiation in commercially available material. Upon annealing, divacancies are created along with other paramagnetic defects. Higher spin species or laser-induced scrambling of the charge states of these paramagnetic impurities may also increase the effect of impurities with respect to CCE.

Furthermore, the observed optical linewidth is significantly broadened by spectral diffusion. In this material, lines are in the 150-350 MHz range. We can use this broadening to estimate[18] the trap density to be 3×10$^{14}$-3×10$^{15}$ cm$^{-3}$ for the *kk* defect, which would be consistent with the observed Hahn echo times if these trap are assumed to be paramagnetic.

*Calculations of coherence functions*

Cluster-correlation expansion (CCE) calculations of the coherence function for the nuclear spins were carried out according to the method outlined by Yang and Liu[59] with the choice of parameters described by Seo *et al.*[35]. We apply the CCE up to second order under the assumption that the flip rate of each pair of electron spins is not impacted significantly by interactions with the spins outside a given pair. The total coherence function ($L$) can be factorized into contributions from electron and nuclear spins, respectively: $L(t) = L_{\text{electron}} L_{\text{nuclear}}$. Further details are found in the Supplemental Information.

*Calculations of nuclear memory availability*

In order to decide whether the nuclear spin at the lattice site $i$ can be used as a memory, we evaluated the state fidelity of the electron spin state after a nuclear induced rotation. The fidelity can be inferred from the electron magnetization along the x-axis. Assuming that nuclei-nuclei interactions are neglegible, the expectation value of the electron magnetization along the x-axis at a given $N$ and $\tau$ in the presence of a nucleus $i$ can be expressed as:

$$\widetilde{M}_i = E(M|M_i \in M) = M_i \prod_{j \neq i} E(M_j) \qquad 1$$

where $M_i$ ($M_j$) is the conditional magnetization when only one nucleus (at lattice site $i$ ($j$)) interacts with the electron, $j$ runs over all other possible nuclear positions, and $E(M_j)$ is the expectation value of the conditional magnetization. A nucleus at lattice site $j$ is considered to be useable as a memory unit if there exist at least one set of $N$ and $\tau$ with $N2\tau$ smaller than a maximum gate time, such that the fidelity of the electron spin after rotation $\widetilde{M}_i(N,\tau)$ is higher than a certain threshold $F_{min}$. The average number of nuclei $i$ present at this lattice site is equal to the concentration of the spin-1/2 isotope $c_i$. The resulting total number of usable memory units is computed as the sum of $c_i$ for all $i$ that meet the fidelity criterion for at least one set of $N, \tau$:

$$N_{\text{mem}} = \sum_i^{F(\widetilde{M}_i) \geq F_{min}} c_i \qquad 2$$

Further details are found in the Supplemental Information.

*Hyperfine cutoff value*

A cutoff of $A_{||}$ = 2π·60 kHz is used in this work as a rough guideline for when hyperfine are low enough to act as optimal quantum memories. This corresponds to hyperfine values that were found to be ideal for communication protocols with the NV⁻ in diamond[8,30]. Coincidentally, this cutoff is also roughly the same order of magnitude as the ODMR linewidth we measure in isotopic samples (20 kHz) and provides an approximate limit for the lowest hyperfine spin which could be considered strongly coupled.

*Errors*

All quoted uncertainties are reported at one standard deviation.

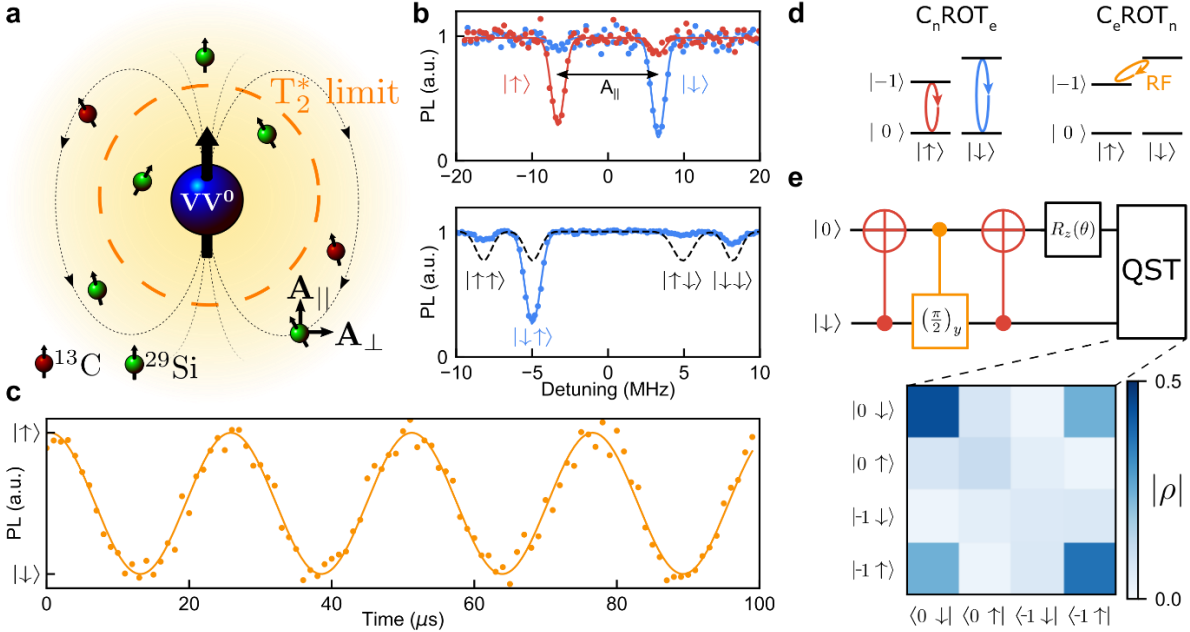

**Fig. 1 | Initializing, controlling and entangling strongly coupled nuclear spins. a,** Schematic of a single divacancy with surrounding nuclear spins. **b,** Optically detected magnetic resonance of a single (*kk*) VV$^0$ after initialization of both the electron and either 1 (top) or 2 (bottom) strongly coupled nuclear spins. Top: initialization in either the $|\uparrow\rangle$ (red) or $|\downarrow\rangle$ (blue) nuclear spin states. Detuning is from 1.139 GHz. Bottom: dashed line (black) represents the expected results from an uninitialized state, blue line is the experimental initialized state. Detuning is from 2.153 GHz. **c,** Nuclear Rabi oscillations (between $|-1\downarrow\rangle$ and $|-1\uparrow\rangle$) obtained by driving an RF tone implementing a C$_e$ROT$_n$. **d,** level structure schematic of a divacancy spin coupled to a single nuclear register. The $|+1\rangle$ electron spin state is not shown. (left) C$_n$ROT$_e$ transitions correspond to the peaks seen in **b**. (right) C$_e$ROT$_n$ RF transition corresponds to the oscillations in **c**. **e,** (top) Quantum circuit used to generate a bipartite entangled state between an electron and nuclear spin. (bottom) Resulting density matrix ($|\rho|$). The third initialized qubit is omitted. All data are taken at T= 5 K.

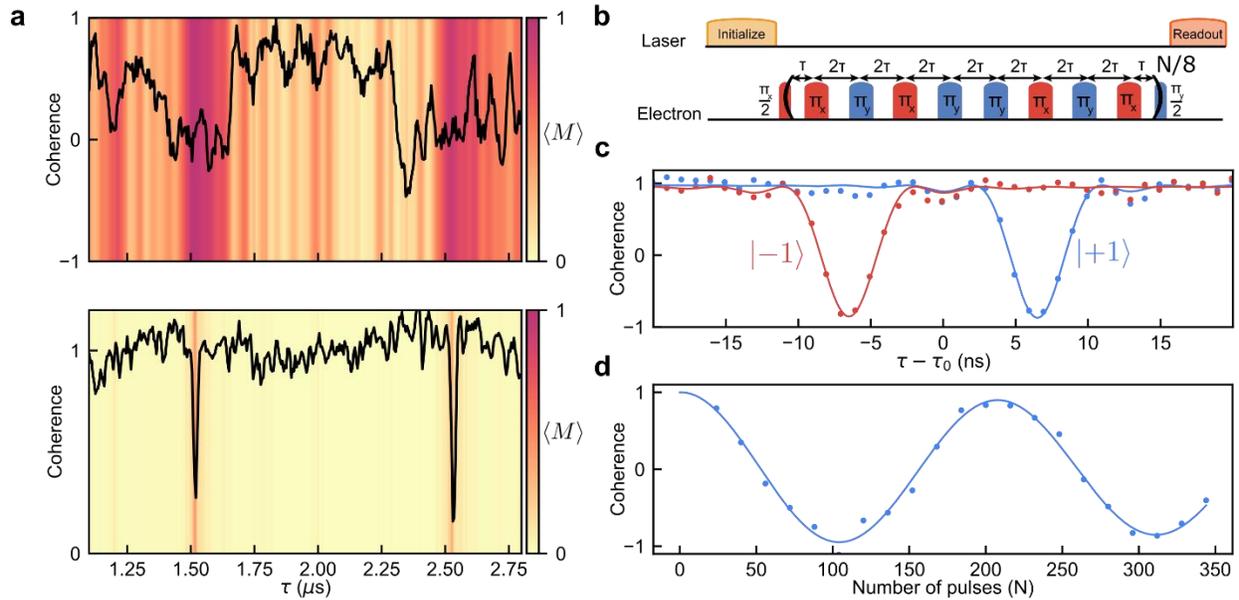

**Fig. 2 | Spectroscopy and control of weakly coupled nuclear spins. a,** CPMG based NMR spectroscopy of the nuclear environment of an example *kk* divacancy in a natural (top) and isotopically purified (bottom) sample. The data is shown as a black solid line. The background gradient represents the calculated average coherence function obtained over many nuclear configurations ⟨$M$⟩, which represents the expected density of coherence dips. **b**, Schematic of the XY8 pulse sequence. **c,** Coherence dips (8$^{th}$ order ($k$=8), $\tau_0$=6.125 μs) using either the $|-1\rangle$ (red) or $|+1\rangle$ (blue) electron spin state, providing a measure of $A_{\parallel} \approx 2\pi \cdot 650$ Hz. **d,** A C$_e$ROT$_{x,n}$(±θ) oscillation demonstrated on the 6$^{th}$ order ($k$=6) of the isolated nuclear spin and achieved by varying the number of XY8 subsequence repetitions. After seven XY8 repetitions (total pulse number $N$=56), a conditional ±π/2 rotation is achieved with a fidelity of F=97(1)%. All data are taken at T= 5 K and B = 584 G.

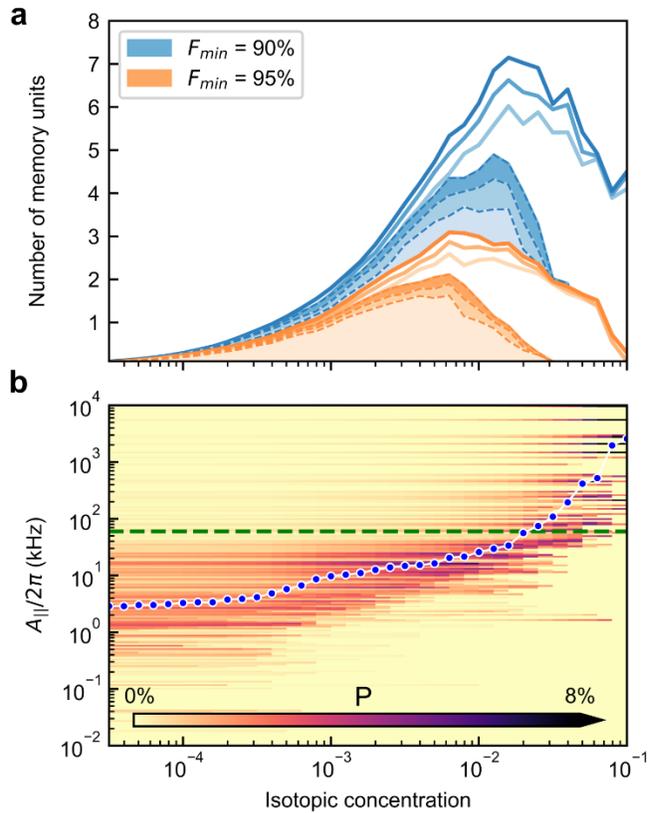

**Fig. 3 | Isotopic optimization of nuclear memories. a**, Calculated average number of memory units as a function of isotopic concentration where [$^{13}$C] = [$^{29}$Si]. A memory unit is defined as a nuclear spin that can be isolated and controlled above a given gate fidelity ($F_{min}$) within the maximum gate time. Solid lines correspond to all memory units whereas the dotted lines with shaded areas correspond to only memories with $A_\parallel <$ 2π·60 kHz. Three different maximum allowable gate times are represented (lightest to darkest: 1 ms, 1.5 ms and 2 ms). **b,** Distribution of the hyperfine values for usable memory units as a function of isotopic concentration. Darker colors correspond to a higher probability (P) that memory units, if present and usable, will have the corresponding hyperfine value (maximum gate time = 1.5 ms, $F_{min}$ = 0.9). Blue circles show the median of the distribution at the given concentration. The green dotted line corresponds to $A_\parallel$ = 2π·60 kHz. The values are computed at the magnetic field of 500 G.

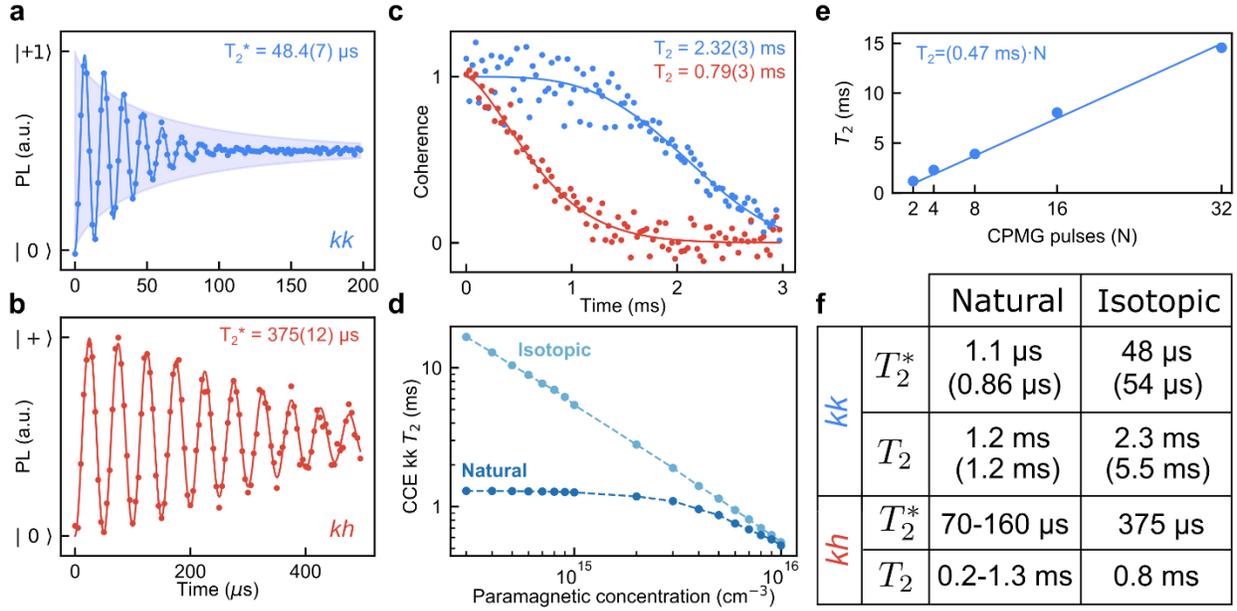

**Fig. 4 | Divacancy dephasing and decoherence times in isotopically purified material. a,** Dephasing of a c-axis (*kk*) defect in the isotopic sample at B=48.8 G. The shaded region represents the predicted average results from CCE (B = 50 G and paramagnetic density of 1×10$^{15}$ cm$^{-3}$). **b,** Dephasing of a basal (*kh*) defect at B = 0 G. **c,** Coherence function under a Hahn echo sequence for *kk* (blue) and *kh* (red) single defects. **d,** CCE calculations (including the effects of paramagnetic traps) for a *kk* defect showing that the expected average Hahn echo T$_2$ varies greatly based on paramagnetic spin density for both natural (dark blue) and isotopic (light blue) material (at B = 500 G). **e,** Coherence time for a (*kk*) defect in the isotopic sample under a varying number of CPMG pulses (*N*) shows that T$_2$ increases roughly linearly with pulse number (B = 48.8 G). **f,** Table summarizing representative numbers for T$_2$* and T$_2$ (Hahn echo) in *kk* and *kh* defects in both natural and isotopic samples. Natural SiC coherences are taken from literature[12,36,37]. Numbers in parentheses are the theoretical numbers obtained by CCE (at B = 50 G) with both the nuclear spin bath and a paramagnetic spin bath of 1×10$^{15}$ cm$^{-3}$. All data are taken at T= 5 K.

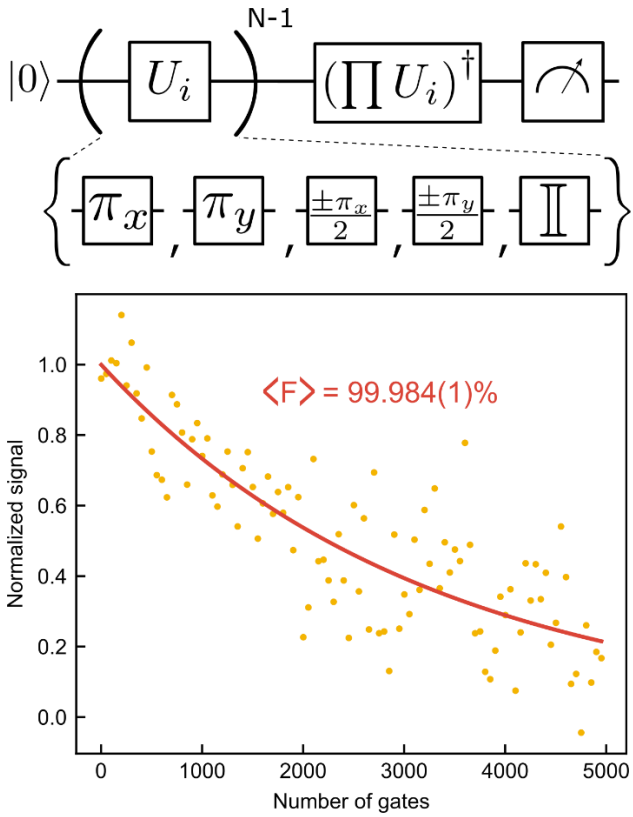

**Fig. 5 | Average single qubit gate fidelity as measured by randomized benchmarking.** Results obtained by applying *N* Clifford gates (as represented by the quantum circuit) on the electronic spin of a *kh* defect in the isotopically purified material at T= 5 K, B= 0 G. From this decay, we extract an average gate fidelity of 99.984(1)%.

**Acknowledgements**

We thank E. O. Glen, S. Bayliss, G. Wolfowicz and P. J. Duda for useful discussions and assistance. We thank *Quantum Opus* for their assistance with SNSPDs. This work made use of the UChicago MRSEC (NSF DMR-1420709) and Pritzker Nanofabrication Facility, which receives support from the SHyNE, a node of the NSF's National Nanotechnology Coordinated Infrastructure (NSF ECCS-1542205). C.P.A., A.B., K.C.M., A.L.C., and D.D.A. were supported by AFOSR FA9550-19-1-0358, DARPA D18AC00015KK1932, and ONR N00014-17-1-3026. T.O. was supported by KAKENHI (18H03770 and 20H00355). J.U.H was supported by the Swedish Energy Agency (43611-1). N.T.S. was supported by the Swedish Research Council (VR 2016-04068) and the Carl Tryggers Stiftelse för Vetenskaplig Forskning (CTS 15:339). J.U.H. and N.T.S. were also supported by the EU H2020 project QuanTELCO (862721) and the Knut and Alice Wallenberg Foundation (KAW 2018.0071).

**Author contributions**

A.B. and C.P.A conceived the experiments, performed the measurements, and analyzed the data. C.P.A fabricated the devices. A.B. and K.C.M developed the experimental setup. M.O, H.M and G.G provided a theoretical framework. M.O. performed the numerical simulations and computations. A.L.C. assisted in device fabrication. H.A. and T.O. performed the electron irradiation. J.U.H. and N.T.S. grew the isotopically purified SiC samples. D.D.A. advised on all efforts. All authors contributed to manuscript preparation.

**Competing interests**

The authors declare no competing financial interests.

**Materials & Correspondence**

Correspondence and requests for materials should be addressed to D.D.A. (awsch@uchicago.edu)

**Data availability**

The data that support the findings of this study are available from the corresponding author upon request.

**References**

1. Hartman, J. S. *et al.* NMR Studies of Nitrogen Doping in the 4H Polytype of Silicon Carbide: Site Assignments and Spin−Lattice Relaxation. *J. Phys. Chem. C* **113**, 15024–15036 (2009).
2. Niedbalski, P. *et al.* Magnetic-Field-Dependent Lifetimes of Hyperpolarized 13C Spins at Cryogenic Temperature. *J. Phys. Chem. B* **122**, 1898–1904 (2018).
3. Terblanche, C. J., Reynhardt, E. C. & Van Wyk, J. A. 13C spin-lattice relaxation in natural diamond: Zeeman relaxation at 4.7 T and 300 K due to fixed paramagnetic nitrogen defects. *Solid State Nucl. Magn. Reson.* **20**, 1–22 (2001).
4. Bradley, C. E. *et al.* A Ten-Qubit Solid-State Spin Register with Quantum Memory up to One Minute. *Phys. Rev. X* **9**, 31045 (2019).
5. Maurer, P. C. *et al.* Room-Temperature Quantum Bit Memory Exceeding One Second. *Science (80-. ).* **336**, 1283–1286 (2012).
6. Zaiser, S. *et al.* Enhancing quantum sensing sensitivity by a quantum memory. *Nat. Commun.* **7**, 1–11 (2016).
7. Taminiau, T. H., Cramer, J., Van Der Sar, T., Dobrovitski, V. V & Hanson, R. Universal control and error correction in multi-qubit spin registers in diamond. *Nat. Nanotechnol.* **9**, 171–176 (2014).
8. Reiserer, A. *et al.* Robust quantum-network memory using decoherence-protected subspaces of nuclear spins. *Phys. Rev. X* **6**, 21040 (2016).
9. Kalb, N. *et al.* Entanglement distillation between solid-state quantum network nodes. *Science (80-. ).* **356**, 928–932 (2017).
10. Cujia, K. S., Boss, J. M., Herb, K., Zopes, J. & Degen, C. L. Tracking the precession of single nuclear spins by weak measurements. *Nature* vol. 571 230–233 (2019).
11. Unden, T. K., Louzon, D., Zwolak, M., Zurek, W. H. & Jelezko, F. Revealing the Emergence of Classicality Using Nitrogen-Vacancy Centers. *Phys. Rev. Lett.* **123**, 140402 (2019).
12. Klimov, P. V, Falk, A. L., Christle, D. J., Dobrovitski, V. V & Awschalom, D. D. Quantum entanglement at ambient conditions in a macroscopic solid-state spin ensemble. *Sci. Adv.* **1**, (2015).

13. Koehl, W. F., Buckley, B. B., Heremans, F. J., Calusine, G. & Awschalom, D. D. Room temperature coherent control of defect spin qubits in silicon carbide. *Nature* **479**, 84–87 (2011).
14. Widmann, M. *et al.* Coherent control of single spins in silicon carbide at room temperature. *Nat. Mater.* **14**, 164–168 (2015).
15. Diler, B. *et al.* Coherent control and high-fidelity readout of chromium ions in commercial silicon carbide. *npj Quantum Inf.* **6**, 1–6 (2020).
16. Wolfowicz, G. *et al.* Vanadium spin qubits as telecom quantum emitters in silicon carbide. *Sci. Adv.* **6**, eaaz1192 (2020).
17. Son, N. T. *et al.* Developing silicon carbide for quantum spintronics. *Appl. Phys. Lett.* (2020) *(Accepted)*
18. Anderson, C. P. *et al.* Electrical and optical control of single spins integrated in scalable semiconductor devices. *Science (80-. ).* **366**, 1225–1230 (2019).
19. Christle, D. J. *et al.* Isolated Spin Qubits in SiC with a High-Fidelity Infrared Spin-to-Photon Interface. *Phys. Rev. X* **7**, 21046 (2017).
20. Crook, A. L. *et al.* Purcell Enhancement of a Single Silicon Carbide Color Center with Coherent Spin Control. *Nano Lett.* (2020) doi:10.1021/acs.nanolett.0c00339.
21. Nagy, R. *et al.* High-fidelity spin and optical control of single silicon-vacancy centres in silicon carbide. *Nat. Commun.* **10**, 1–8 (2019).
22. Taminiau, T. H. *et al.* Detection and control of individual nuclear spins using a weakly coupled electron spin. *Phys. Rev. Lett.* **109**, 137602 (2012).
23. Rong, X. *et al.* Experimental fault-tolerant universal quantum gates with solid-state spins under ambient conditions. *Nat. Commun.* **6**, 1–7 (2015).
24. Son, N. T. *et al.* Divacancy in 4H-SiC. *Phys. Rev. Lett.* **96**, 055501 (2006).
25. Waldherr, G. *et al.* Quantum error correction in a solid-state hybrid spin register. *Nature* **506**, 204–207 (2014).
26. Falk, A. L. *et al.* Optical Polarization of Nuclear Spins in Silicon Carbide. *Phys. Rev. Lett.* **114**, 247603 (2015).
27. Fuchs, G. D., Burkard, G., Klimov, P. V & Awschalom, D. D. A quantum memory intrinsic to single nitrogen-vacancy centres in diamond. *Nat. Phys.* **7**, 789–793 (2011).

28. Pfaff, W. *et al.* Demonstration of entanglement-by-measurement of solid-state qubits. *Nat. Phys.* **9**, 29–33 (2013).
29. Gurudev Dutt, M. V *et al.* Quantum register based on individual electronic and nuclear spin qubits in diamond. *Science (80-. ).* **316**, 1312–1316 (2007).
30. Kalb, N., Humphreys, P. C., Slim, J. J. & Hanson, R. Dephasing mechanisms of diamond-based nuclear-spin memories for quantum networks. *Phys. Rev. A* **97**, 062330 (2018).
31. Hensen, B. *et al.* Loophole-free Bell inequality violation using electron spins separated by 1.3 kilometres. *Nature* **526**, 682–686 (2015).
32. Aslam, N. *et al.* Nanoscale nuclear magnetic resonance with chemical resolution. *Science (80-. ).* **357**, 67–71 (2017).
33. Müller, C. *et al.* Nuclear magnetic resonance spectroscopy with single spin sensitivity. *Nat. Commun.* **5**, 1–6 (2014).
34. Abobeih, M. H. *et al.* Atomic-scale imaging of a 27-nuclear-spin cluster using a quantum sensor. *Nature* **576**, 411–415 (2019).
35. Seo, H. *et al.* Quantum decoherence dynamics of divacancy spins in silicon carbide. *Nat. Commun.* **7**, 1–9 (2016).
36. Miao, K. C. *et al.* Electrically driven optical interferometry with spins in silicon carbide. *Sci. Adv.* **5**, (2019).
37. Miao, K. C. *et al.* Universal coherence protection in a solid-state spin qubit. *In preparation* (2020).
38. Christle, D. J. *et al.* Isolated electron spins in silicon carbide with millisecond coherence times. *Nat. Mater.* **14**, 160–163 (2015).
39. Bonato, C. *et al.* Optimized quantum sensing with a single electron spin using real-time adaptive measurements. *Nat. Nanotechnol.* **11**, 247–252 (2016).
40. Bauch, E. *et al.* Ultralong Dephasing Times in Solid-State Spin Ensembles via Quantum Control. *Phys. Rev. X* **8**, 31025 (2018).
41. Ishikawa, T. *et al.* Optical and Spin Coherence Properties of Nitrogen-Vacancy Centers Placed in a 100 nm Thick Isotopically Purified Diamond Layer. *Nano Lett* **12**, 34 (2012).
42. Clerk, A. A., Lehnert, K. W., Bertet, P., Petta, J. R. & Nakamura, Y. Hybrid

quantum systems with circuit quantum electrodynamics. *Nat. Phys.* 1–11 (2020) doi:10.1038/s41567-020-0797-9.

43. Whiteley, S. J. *et al.* Spin–phonon interactions in silicon carbide addressed by Gaussian acoustics. *Nat. Phys.* **15**, 490–495 (2019).

44. Jamonneau, P. *et al.* Competition between electric field and magnetic field noise in the decoherence of a single spin in diamond. *Phys. Rev. B* **93**, 024305 (2016).

45. Balasubramanian, G. *et al.* Ultralong spin coherence time in isotopically engineered diamond. *Nat. Mater.* **8**, 383–387 (2009).

46. Herbschleb, E. D. *et al.* Ultra-long coherence times amongst room-temperature solid-state spins. *Nat. Commun.* **10**, 1–6 (2019).

47. Witzel, W. M., Carroll, M. S., Cywiński, A. & Das Sarma, S. Quantum decoherence of the central spin in a sparse system of dipolar coupled spins. *Phys. Rev. B - Condens. Matter Mater. Phys.* **86**, 035452 (2012).

48. Bar-Gill, N. *et al.* Suppression of spin-bath dynamics for improved coherence of multi-spin-qubit systems. *Nat. Commun.* **3**, 1–6 (2012).

49. Yang, C. H. *et al.* Silicon qubit fidelities approaching incoherent noise limits via pulse engineering. *Nat. Electron.* **2**, 151–158 (2019).

50. Yoneda, J. *et al.* A quantum-dot spin qubit with coherence limited by charge noise and fidelity higher than 99.9%. *Nat. Nanotechnol.* **13**, 102–106 (2018).

51. Knill, E. Quantum computing with realistically noisy devices. *Nature* **434**, 39–44 (2005).

52. Veldhorst, M. *et al.* An addressable quantum dot qubit with fault-tolerant control-fidelity. *Nat. Nanotechnol.* **9**, 981–985 (2014).

53. Campbell, E. T., Terhal, B. M. & Vuillot, C. Roads towards fault-tolerant universal quantum computation. *Nature* vol. 549 172–179 (2017).

54. Petersen, E. S. *et al.* Nuclear spin decoherence of neutral 31 P donors in silicon: Effect of environmental 29 Si nuclei. *RAPID Commun. Phys. Rev. B* **93**, 161202 (2016).

55. Guichard, R., Balian, S. J., Wolfowicz, G., Mortemousque, P. A. & Monteiro, T. S. Decoherence of nuclear spins in the frozen core of an electron spin. *Phys. Rev. B* **91**, 214303 (2015).

56. Dong, W., Calderon-Vargas, F. A. & Economou, S. E. Precise high-fidelity electron-nuclear spin entangling gates in NV centers via hybrid dynamical decoupling sequences. *arXiv:2002.01480* (2020).
57. Oliveros, A., Guiseppi-Elie, A. & Saddow, S. E. Silicon carbide: A versatile material for biosensor applications. *Biomed. Microdevices* **15**, 353–368 (2013).
58. Lebedev, A. A. *Radiation Effects in SiC*. vol. 6 (Materials research forum, 2017).
59. Yang, W. & Liu, R. B. Quantum many-body theory of qubit decoherence in a finite-size spin bath. *Phys. Rev. B - Condens. Matter Mater. Phys.* **78**, 085315 (2008).